# Comparative study of photo-induced electronic transport along ferroelectric domain walls in lithium niobate single crystals

L. L. Ding,[1,2,3,4,a)] E. Beyreuther,[2] B. Koppitz,[2] K. Kempf,[2] J. H. Ren,[1,3,4] W. J. Chen,[1,3,4,5] M. Rüsing,[2,6] Y. Zheng,[1,3,4] and L. M. Eng,[2,7,a)]

[1]Guangdong Provincial Key Laboratory of Magnetoelectric Physics and Devices, School of Physics, Sun Yat-sen University, Guangzhou 510275, China

[2]Institute of Applied Physics, TUD Dresden University of Technology, 01062 Dresden, Germany

[3]State Key Laboratory of Optoelectronic Materials and Technologies, School of Physics, Sun Yat-sen University, Guangzhou 510275, China

[4]Centre for Physical Mechanics and Biophysics, School of Physics, Sun Yat-sen University, Guangzhou 510275, China

[5]School of Materials, Shenzhen Campus of Sun Yat-sen University, Shenzhen 518107, China

[6]Institute of Photonic Quantum Systems, Integrated Quantum Optics, Paderborn University, 33098 Paderborn, Germany

[7]ct.qmat: Dresden-Würzburg Cluster of Excellence–EXC 2147, TU Dresden, 01062 Dresden, Germany

[a)]Authors to whom correspondence should be addressed: lukas.eng@tu-dresden.de, dinglli@mail2.sysu.edu.cn

**Abstract**

Ferroelectric domain wall conductivity (DWC) is an intriguing and promising functional property, that can be elegantly controlled and steered through a manyfold of external stimuli such as electric and mechanical fields. Optical-field control, as a non-invasive and flexible handle, has rarely been applied so far, but significantly expands the possibility for both tuning and probing DWC. On the one hand, as known from Second-Harmonic, Raman, and CARS micro-spectroscopy, the optical in-and-out approach delivers delicate parameters on the DW distribution, the DW inclination, and probes the DW vibrational modes; on the other hand, photons might be applied also to directly generate charge carriers within the DW as well, hence acting as a functional and spectrally tunable probe to deduce the integral or local absorption properties and bandgaps of conductive DWs. Here, we report on such an optoelectronic approach by investigating the photo-induced DWC (PI-DWC) in DWs of the single-crystalline model system lithium niobate, a material that is well known for hosting conductive DWs. We compare three different crystals containing a very different number of domain walls, namely: (A) none, (B) one, and (C) many conductive DWs. All three samples are inspected for their current-voltage (I-V) behavior (i) in darkness, and (ii) for different illumination wavelengths swept from 500 nm down to 310 nm. All samples show their maximum PI-DWC at 310 nm, i.e., at the optical bandgap of lithium niobate; moreover, sample (C) reaches PI-DWCs of several µA. Interestingly, a noticeable PI-DWC is also observed for sub-bandgap illumination, i.e., wavelengths as high as 500 nm, hinting towards the existence and decisive role of electronic in-gap states that contribute to the electronic charge transport along DWs. Finally, complementary conductive atomic force microscopy (c-AFM) investigations under illumination proved that the PI-DWC indeed is confined to the DW area only, and does not originate from any photo-induced bulk conductivity.

**Main Text**

Ferroelectric materials exhibit a spontaneous and stable dielectric polarization, resulting in a rich and variable assembly of domain and domain wall (DW) structures that is receiving continued attention.[1-3] Furthermore, the multifield-controllable electrical transport across these structures offers many prospects for the vivid application of ferroelectrics into electronic devices, such as ferroelectric sensors, memristors, and even ferroelectric synaptic circuits.[4-10] Since the conductivity characteristic of DWs in general is a decisive performance figure to evaluate the functionality of any electronic component, it is significant to clarify the electrical transport properties of ferroelectric materials with polar microstructures in detail.[11-13] For more than a decade, enhanced electrical-conductivity phenomena in ferroelectric DWs have been reported in several typical ferroelectric materials, such as $BiFeO_3$ (BFO), $Pb(Zr_{0.2}Ti_{0.8})O_3$ (PZT), or $LiNbO_3$ (LNO), where the first artificial enhanced electric vortex core, a large a.c. conductance of uncharged DW, or an inclination-tunable charged-DW conductivity were found, respectively.[14-17] These DW-related abundant polar states make it possible to regulate and exploit the conductance at the nanometer length scale in a confined manner. Considering the influence of several main factors on DW formation, for instance, the lattice structure of the ferroelectric material, domain-growth conditions and manufacturing parameters,[18-21] the selection of the ferroelectric material itself as well as the adjustment of DW shape, structure, and density, will all affect the DW conductivity and thus offer many directions for device design and development.[22,23]

DWs constitute the interface between different orientations of the order parameter, which - on the lattice scale - will provoke many changes in numerous physical factors such as charge carrier density, the (dielectric) polarization, or the oxygen vacancy density.[24-26] Several novel domain structures of, for example, $PbTiO_3$ and $BiFeO_3$ have been discovered so far, like vortex, four-quadrant central divergent, convergent, cross-shape domains etc.,[27-29] and their DWs exhibit an excellent and controllable electrical transport behavior. An example of charged domain walls (in BFO center-type domain) has been confirmed for a conductance difference of three orders of magnitude under domain pattern switching.[28] Applying electrical and mechanical fields as external control parameters are the common methods to modulate the domain wall morphology and its transport properties. Whereas, in the field of classical semiconductors' light dependent transport properties, i.e., photoconductivity, plays a central role and forms the foundation of many optoelectronic applications, such as light sensors, detectors, or imaging devices. In contrast, the photo-transport properties of (non-)conductive ferroelectric domain walls remain less studied so far. On the one hand this may form the basis of DW-based opto-electronic devices, but at the same time also provide a fundamental insight into the microscopic mechanisms of photo-induced transport, e.g., relevant in-gap energy band levels, which are also not well understood.

Single crystals of lithium niobate ($LiNbO_3$, LNO) which can be considered as a model ferroelectric, are widely available due to its well-known use for non-linear optical, piezoelectric, and pyroelectric applications.[30] LNO is an uniaxial ferroelectric in which the polarization points only along two opposite orientations, i.e., along the ± z-axes. Therefore, under equilibrium, we speak of 180°-domains and associated 180°-domain walls (180°-DWs). Due to the lattice structure of LNO, domains typically show a hexagonal structure.[30] In the absence of any dopants or DWs, bulk LNO exhibits an optical bandgap of around 4 eV and thus constitutes an archetypical electrical insulator with a bulk electrical conductivity of $10^{-18}$ (mΩ cm)$^{-1}$ at room temperature.[17,31] Nonetheless, engineered DWs may increase that conductivity by 7 orders of magnitude or more confined to a cross-section of nanometers or less.[17,32] Hence, in the work presented here, we investigate the DWC in 5mol% MgO-doped LNO single crystals under different illumination conditions, and moreover show that solely these DWs contribute to the photo-induced domain wall current (PI-DWC).

Domains as needed for our experiments are fabricated into commercially available, monodomain 5mol%-MgO-doped LNO single crystals (z-cut, 200 μm thickness, Yamaju Ceramics, Inc., Japan) using UV-assisted electric-field poling with liquid electrodes,[33] as schematically depicted in Fig. 1(a). Freshly poled inverted domains typically arrange with their order parameter aligned parallel to the spontaneous polarization and are not conductive, hence forming noncharged 180°-DWs. Applying the high-voltage method of Godau et al. induces conductive DWs, which we refer to as "enhanced".[32] More concretely, after inducing a DW, we vapor-deposit chromium electrodes (1x1 mm$^2$) onto the two ± z-crystal faces of our samples (5x6 mm$^2$) such that the freshly generated DWs are fully covered, and then apply a high voltage of up to 500 V (limited to 1/3 of the coercive field) in order to obtain conductive (tilted) DWs, as shown in Fig. 1(b) (I-V curves as obtained before applying this "enhancement" process are shown in the Supplementary Fig. S1). To further enhance both the conductivity and the density of domains, we apply a larger voltage of up to 900 V on the basis of one conductive domain wall, the voltage ramping rate is 10 V/s, so the voltage applies for a up to 90 s [see Fig. 1(c)]. During this procedure, a large portion of the originally poled domain fractures into many sub-domains. For further discussion, we label this process as the "super-enhancement" process, which provides a high density of DWs [see sample (C)] and a larger total DWC. We control and visualize the domain and DW distribution after every step by applying second harmonic generation (SHG) microscopy (for details see Supplementary S2);[34] the initial domain configuration in Fig. 1(d) presents a single, hexagonally shaped domain, while after super-enhancement, a large number of stripe-like domains has formed across the single crystal [see Fig. 1(e)]. SHG imaging then was carried out in 3D to obtain the DWs depth distribution and alignment. From both the kinetics of poling and our previous works,[32,34] it is obvious that the domains shrink in their size as a function of depth [Fig. 1(f)], resulting in a significant DW inclination of 1.12° with respect to the crystallographic c-axis after super-enhancement.

In order to quantitatively study the photo-induced electrical transport properties of the DWs, we compare the current-voltage characteristics of three LNO samples of different DW density, which are depicted in Fig. 2: (A) a monodomain sample, (B) a sample containing exactly one single poled domain only (after conductivity enhancement, hexagon diameter around 190 μm), and (C) a multiple-domain sample (after "super-enhancement"), which are labeled as sample (A), (B), and (C), respectively. All samples were contacted to an electrometer (Keithley 6517B) using the same sized Cr electrodes (1×1 mm$^2$) as described above for the "enhancement protocol". The Cr electrodes are only 5 nm thin, and thus can be considered as being quasi optically transparent[35] for our photo-induced domain wall conductivity (PI-DWC) experiments (cf. transmittance values under different wavelength illumination conditions, which are shown in Supplementary Table S1). Therefore, the spectral range was filtered for excitation into five discrete center wavelengths (500 nm, 400 nm, 350 nm, 320 nm, and 310 nm) always illuminating the samples' +z surfaces with a constant photon flux of 9.5×10$^{13}$ s$^{-1}$; note that the latter two selected wavelengths correspond to photon energies slightly below and above the LNO bandgap energy. The illumination setup mainly consists of a 1000 W Xenon arc lamp, a grating monochromator which has a spectral bandwidth of about 10 nm, an edge filter as well as a neutral density filter to control the photon flux, and two pairs of the fused-silica lenses, that will finally focus the beam onto the sample surface (the spot diameter is around 5 mm).[36] We recorded current-voltage (I-V) characteristics of the three samples (i) in darkness and (ii) at these five different illumination wavelengths. Here, the voltage was swept in 1-V increments per 2 seconds from -10 V → 0 V, then from 0 V → +10 V, and finally from 10 V back to -10 V. The current was measured by a Keithley 6517B electrometer which can measure currents over 15 orders of magnitude from 10 aA up to 20 mA. We captured three cycles to rule out possible instabilities and artefacts.

For sample (A) the monodomain reference sample, the maximum current in darkness measures approximately $5.68 \times 10^{-12}$ A at a +10 V bias voltage, which is a very low current but as expected since LNO is a wide-bandgap insulator in its ground state. The current changes when illuminating sample (A) with the five different wavelengths, as presented in Fig. 2(a). In particular, the currents under illumination with 500 nm, 400 nm, and 350 nm wavelengths show PI-DWC values very close to the dark current on the order of a few pA at ± 10 V [the zoomed-in I-V curve in darkness and under the wavelengths of 500 nm, 400 nm, 350 nm is shown in Fig. S2(a)]. However, when using the 320 nm and 310 nm photons, we observe a large PI-DWC increase by up to 3 orders of magnitude, reaching an overall maximum current threshold of $2.21 \times 10^{-9}$ A for the above-mentioned photon flux. This photocurrent stems from the bulk photoconductivity of sample (A) since it does not contain any domain walls.

Next, we apply the same measurement procedure to samples (B) and (C). These two samples both exhibit a non-linear diode-like I-V characteristics in general: when a positive voltage is applied, the current is "on", while under negative measuring voltages there is nearly no current flowing through the DW(s). As we know, metal-DW interfaces play an important role in these transport characteristics.[37] After the application of the conductivity-enhancement protocol, the larger near-surface inclination angles of the DWs lead to an improved charge carrier injection from the metal electrode into the DW, or *vice versa*. This kind of local symmetry breaking may severely influence and alter the I-V curve characteristics.[32] In previous studies, these kinds of phenomena and possible origins of the different I-V curves have been studied.[37] As depicted in Figs. 2(b) and (c), sample (B) shows that the DWC at +10 V in the dark measures around 0.6 μA, and sample (C) has already reached more than 5 μA. Hence, to clearly display the changes in the PI-DWC for samples (B) and (C), we also show close-up view of the I-V curves (the voltage varies from + 8.5 V to + 10 V, only) in Figs. S2(c) and (d). It is obvious that the DWC of both samples (B) and (C) increases with decreasing wavelength; moreover, due to the large DW density in sample (C), we observe here a distinctly increased PI-DWC of up to ~7 μA. In order to better compare the currents measured in samples (B) and (C) that obviously flow through very different cross sections, we calculate the "maximum-current ratio" $I_{ratio}$ which is equal to $I_{max}$ (the current value at the +10 V measuring voltage and under the illumination wavelength of 310 nm) divided by $I_{dark}$. Then we define the photoconductance value "*PC*" as the percentage of current increase on basis of $I_{ratio}$ [the details are shown in Table S2, so *PC* = ($I_{ratio}$ - 1)×100% ]. The *PC* of sample (C) measures 30%, which is higher than that of sample (B), having a *PC* of 20% only, indicating that the high-density domain walls indeed show a clearly higher photoconductivity effect. Both samples (B) and (C) reach their maximum currents at 310-nm illumination, demonstrating, as expected, that the electrons in the valence band effectively absorb the photon energy and are excited into the conduction band, and thus generate electron-hole pairs under the light excitation corresponding to the wavelength of the LNO bandgap.

To investigate the temporal (transient) behavior of the photoconductance of these three samples under variable illumination conditions, we recorded the time-dependent PI-DWC at a constant sample bias voltage of +10 V. The experimental procedure is as follows: After mounting the sample, it is first kept in complete darkness for one day using a shutter, diminishing the effects of former illumination, and allowing the DWs to settle down into their ground state. Then we open the shutter for 300 s, and light of the desired wavelength can excite the DWs. Lastly, we close the shutter again and continue recording the current for another 400 s, now again in the dark.

Here, for better comparability of the transients at the respective wavelengths, the normalized photoconductance *PC*$_{norm}$ is employed,[36] which is defined as follows:

$$PC_{norm} = (\frac{I-I_{dark}}{I_{dark}}) \times 100\% = (\frac{I}{I_{dark}} - 1) \times 100\%.$$

In this formula, *I* represents the current that we measure under different illumination wavelengths, while $I_{dark}$ is the current in darkness. Figs. 2(d), (e) and (f) exhibit the normalized *PC*$_{norm}$ values of samples (A), (B), and (C) for different time intervals – we refer to the ascending parts (after switching the illumination on) as "light-on" transients, and the descending parts (after blocking the light) as "light-off" transients. We find the light-on and light-off transients of the *PC*$_{norm}$ to be dramatically different. For sample (A) and sub-bandgap wavelengths (i.e., 500 nm, 400 nm, and 350 nm), the photocurrent transients show the weakest effect [the zoomed-in photocurrent transients under the sub-bandgap wavelengths of sample (A) are shown in Fig. S2(b)], while a fast, nearly stepwise, increased DWC is visible at a 320 nm and 310 nm wavelengths, consistent again with bridging the optical bandgap close to 310 nm. However, the absolute current values (in the range of a few nA) are still relatively low - especially when compared to samples (B) and (C) [the raw-data current changes over time in darkness and under different illumination conditions for the three samples are shown in Fig. S3, where we clearly see that the PI-DWC of samples (B) and (C) is more than three orders of magnitude larger than that of sample (A)]. Compared to sample (B), sample (C) also shows a higher *PC*$_{norm}$ value, which means that the high-density DWs might generate a higher normalized current ratio under light excitation as compared to the dark [see Figs. 2(e) and (f)]. Different to sample (A), the photocurrents of samples (B) and (C) show a slower saturation behavior dependent on time, and there is a long relaxation time even for several hours after the shutter has been closed. In order to extract the characteristic relaxation times of samples (B) and (C), we use a stretched-exponential decay model in the form $I_{norm}(t) = A \cdot exp[-(t/\tau)^\beta] + I_{norm,\infty}$, with $I_{norm}(t)$ being the normalized photocurrent at the time t, $\tau$ being a characteristic time constant for the underlying process, and $I_{norm,\infty}$ being the normalized final current for t → ∞. The fitting details and $\tau$ as a function of wavelength for samples (B) and (C) are shown in Figs. S4(a) and (b). For sample (B), we could only fit the light-off period, while for sample (C), both light-on and light-off periods can be fitted well. Then the different fitting $\tau$ values of these two samples are plotted in Fig. S4(c). First, we see there is a typical order of magnitude difference of the ascending and descending periods. In the ascending period of sample (C) the $\tau$ values are relatively smaller for wavelengths at 310 nm and 320 nm in comparison to the sub-bandgap wavelengths (500 nm, 400 nm, and 350 nm), which means that the illumination close to the bandgap energy can induce the DW current to reach saturation more quickly. Whereas in both descending periods, most $\tau$ values fitted from the wavelengths of 310 nm, 320 nm, and 350 nm, are larger than that from 500 nm, besides the case of 400 nm in sample (C). This might be the case because more electron-hole pairs have been generated and diffuse under the illumination of the near-bandgap excitation, then after illumination, these free electrons are re-captured to form bound polarons during relaxation, thus it takes longer to reach the original stable state. Especially the case of sample (C) after a 310 nm-wavelength optical excitation, shows the maximum relaxation time in the near-bandgap range. Sample (C) exhibits a high density of charged DWs, hence, more electrons concentrate in the DW area and jump to the conduction band upon illumination, although its activation energy might generally be higher than in the single domain sample (typical activation energies of a single-domain sample and a multi-domain sample, obtained from temperature-dependent DW-current measurements, are given in Table S3). Another speculation about this large $\tau$ might be that due to the formation of a large number of DWs in sample (C), the first poling process and the super-enhancement process may be accompanied with the formation of a decisive concentration of bound polarons or other defects.[38] Thus the barrier increase will make neutralizing of bound polarons become harder, and therefore, increase the photogenerated-electron life time, leading to a long relaxation time constant. Thus, we could use the high-density DW, like in sample (C), to tune the relaxation time as one of the applications in photoelectric device.

Interestingly, as we know, the bandgap of LNO is around 310 nm, but we observe also a significant photoconductance far above this wavelength, for instance, at 500 nm and 400 nm. Considering the spectral absorption cross sections by polarons in Mg-doped congruent LNO, the attribution of the "2.5 eV absorption peak" was explained by transition from a bi-polaron state to the conduction band,[39] which might play a role here as well. Furthermore, a visible luminescence band in LNO has been reported in the past - there is a strong dependence on Mg-doping and the recombination of free electrons with free hole polarons. In particular, in former works, 5mol% Mg-doped LNO showed the relative higher spectral intensity at the energy of around 2.6 eV, with that peak being particularly broad at room temperature.[40] Therefore, the conductivity, dependent on the near bandgap and sub-bandgap illumination wavelength, can improve the understanding of the physical mechanism of the interaction among photon, electron, and polaron; meanwhile this demonstrates the regulatory role of high density DWs on the strength of photoconductivity and its relaxation time.

In order to confirm that the current in the illumination case is indeed confined to the DW areas only and not stemming from the LNO bulk, we use a scanning probe microscope (SPM, NX10 setup from Park

Systems Corp.) to characterize the PI-DWC at the nanoscopic scale. In Fig. 3(a), a schematic drawing of the SPM setup equipped with a set of light-emitting diodes (LEDs) (330 nm, 310 nm, Roithner Lasertechnik GmbH) is depicted. Fig. 3(b) shows the piezoresponse force microscopy (PFM) scanning image of sample (C) displaying the details of the DW arrangement at the sample surface. In these experiments, we use an aluminum reflex coating tip with a free mechanical resonance frequency of 43.5 kHz. Then we perform conducting atomic force microscopy (c-AFM) of the same sample surface with the results been shown in Figs. 3(c), (d), (e). They demonstrate, as expected, that the currents are confined to the DWs of sample (C), even in darkness [Fig. 3(c)]. Next, we show the c-AFM images in Fig. 3(d) and (e) that were scanned under a 330 nm and a 310 nm illumination, respectively, utilizing a power meter (PM 100, Thorlabs GmbH) to roughly keep the photon flux at the same level as was used for the macroscopic integral measurements above (optical output power of around 0.9 mW). From these c-AFM images, we see that DW areas show an obvious distinction and the conductivity at the DWs' site intensifies as the wavelength is lowered. For describing this phenomenon more quantitatively, we analyze the c-AFM results by superimposing the PFM image (i.e., the map of domain distribution) to each c-AFM image separately. As a result, we are now able to precisely extract where the DWC stems from, whether from the DW directly, or any adjacent bulk area. Figs. 3(f), (g) and (h) show the current distribution histograms in darkness and under the two illumination wavelengths. Two different colors are used to represent these two areas. Each vertical bar represents DWC counts in the corresponding c-AFM image. According to the histogram results, all higher-current regions originate from the DW area; notably when illuminating at 310 nm, the histogram exhibits taller-current bars at the DW position, as compared to both the dark and the 330 nm-wavelength illumination. These results hence directly prove that the PI-DWC majorly stems from the conductive DWs in our LNO single crystals, and not from the LNO bulk.

In summary, we demonstrated that the domain wall current in inclined DWs of 5mol%-Mg-doped LNO can be significantly tuned by both super- and sub-bandgap illumination. Whenever there are DWs in the sample, the photo-induced domain wall conductivity (PI-DWC) is confined to the DW area. Only for near-bandgap photo-excitation at ~310 nm a significant bulk photo-current contribution is observed, as expected. Most interestingly, PI-DWC is also measured for sub-bandgap excitation, indicating that electronic trap states within the DW must contribute to the overall DWC mechanism. This promises to incorporate conductive DWs in LNO for optoelectronic applications even below the optical bandgap. Our results will promote optical-field modulation as one possible way for DW configuration and conductance engineering.

See the supplementary material for additional current-voltage data, details on the second-harmonic-generation (SHG) microscopy experiments, plots and curve fitting of photocurrent transients, as well as activation energies.

**Acknowledgment**

We acknowledge financial support from the International Postdoctoral Exchange Fellowship Program through the Guangdong Province, China, the Postdoc Starter Kit Funding Program provided by the Technische Universität Dresden, Germany, the Grants from National Natural Science Foundation of China (No. 12132020), the Guangdong Provincial Key Laboratory of Magnetoelectric Physics and Devices (No. 2022B1212010008), the Deutsche Forschungsgemeinschaft (DFG) through CRC1415 (ID: 417590517), the FOR5044 (ID: 426703838), as well as through the Dresden-Würzburg Cluster of Excellence on "Complexity and Topology in Quantum Matter"—ct.qmat (EXC 2147, ID: 39085490). This work was supported by the Light Microscopy Facility, a Core Facility of the CMCB Technology Platform at TU Dresden.

**Data Availability**

The data that support the findings of this study are available from the corresponding authors upon reasonable request.

**Figures & Captions**

Figure 1

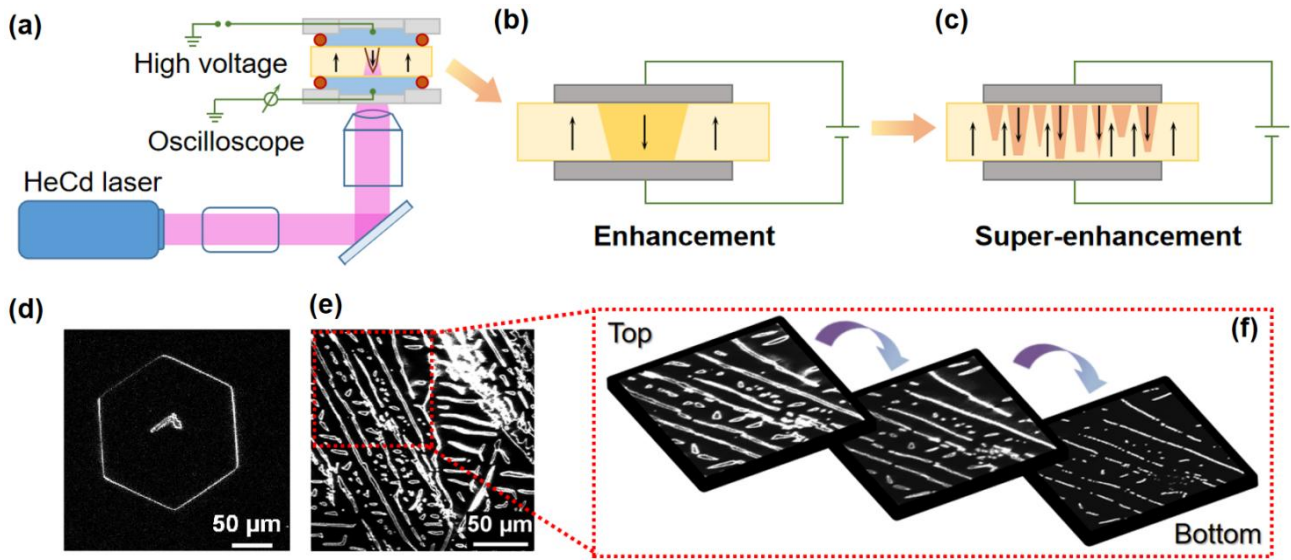

Figure 1:

Generation and visualization of high-density domain walls in LiNbO$_3$ single crystals: (a) UV-assisted poling setup using a liquid cell[33] in order to induce one single hexagonally-shaped domain, as schematically depicted in (b). Note that the domain wall (DW) is charged and conducting when following the standard DW enhancement process.[32] (c) The enhancement process was extended here by applying larger electrical fields (close to the coercive field) for longer time periods, hence allowing for domain switching. As a result of such a "super enhancement" process, we obtain an array of high density DWs. (d, e) Second Harmonic Generation (SHG) microscopy[34] is applied to image the DW distributions of samples (B) and (C), as displayed in (d) and (e), respectively. (f) Zoomed-in pictures and 3D-SHG imaging of the high-density DW sample (C) with the three SHG images being taken at different depth into the crystal, i.e., from the top surface to the middle to the back/bottom interface. Note the gradual shrinking of these domain regions, that hence results in charged DWs.

Figure 2

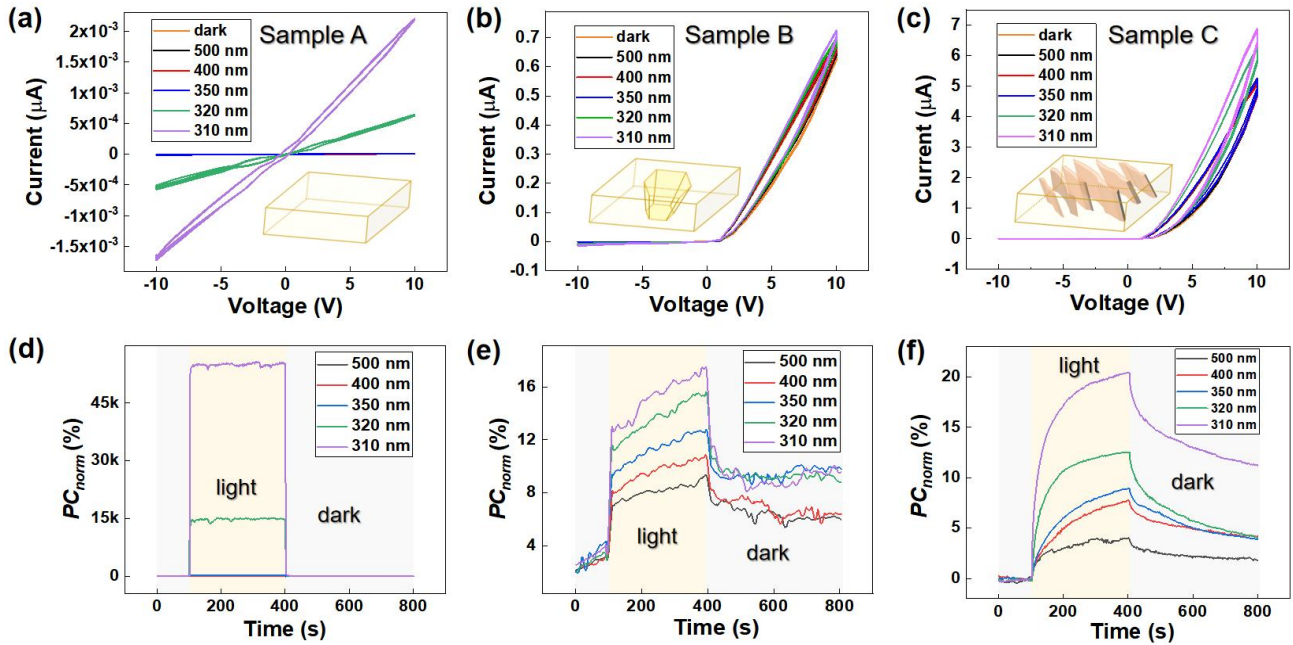

Figure 2:

(a) to (c): Domain-wall-current (DWC) vs. voltage (I-V) curves recorded both in the dark as well as under a constant photon flux at wavelengths ranging between 500 nm and 310 nm. Displayed are I-V curves recorded between ± 10 V for three different DW density samples in LNO, (a) a monodomain sample [no DWs at all, sample (A)], (b) one poled domain only [standard enhancement, see Fig. 1(d), sample (B)], and (c) a multiple domain sample, i.e., after "super-enhancement" [see Fig. 1(e), sample (C)], respectively. Note the very low DWC in (a) in darkness, as is expected for this wide-bandgap semiconductor ($E_g$ = ~4 eV). Nonetheless, the DWC increases heavily for near bandgap illumination at 310 nm, reaching a value of ~2 nA at + 10 V. Notably, the DWCs are 1000 x and 10.000 x larger in (b) and (c), and moreover, show a diode-like I-V behavior. (d) to (f): The normalized photoconductance $PC_{norm}$ for the three different samples (A), (B), (C) under a +10 V bias voltage. After recording the DWC in darkness, light is turned on for 300 s, then switched off for another 400 s. All three samples show the maximum current for a 310 nm (bandgap) illumination.

Figure 3

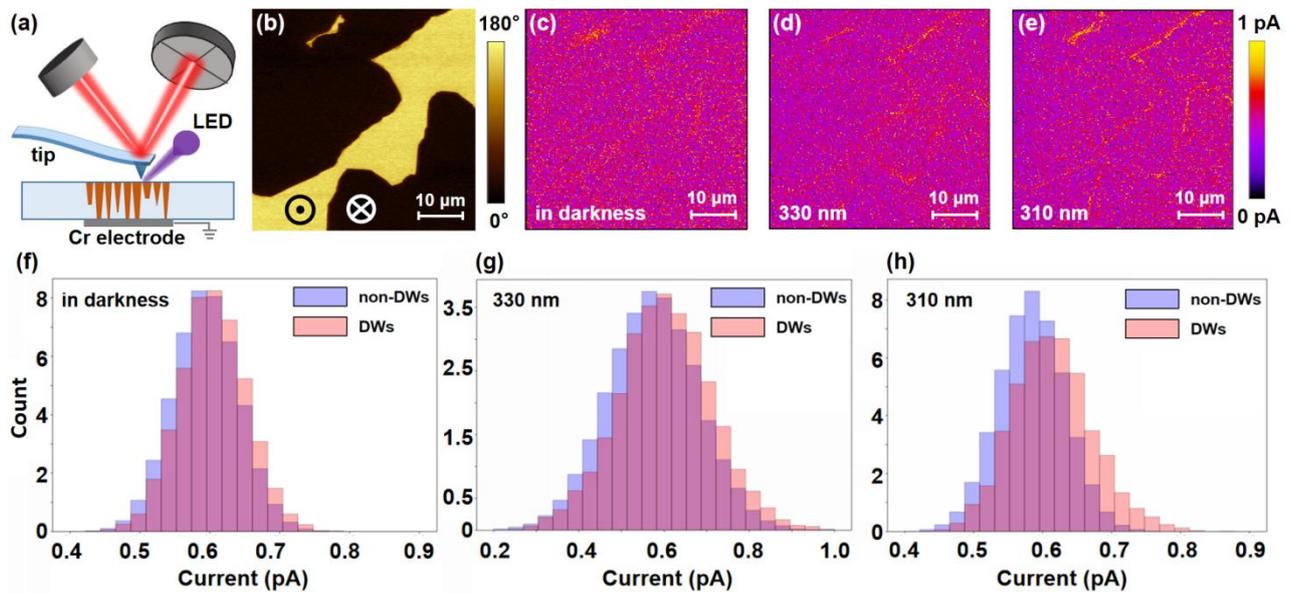

Figure 3:

Conducting atomic force microscopy (c-AFM) measurement of the high-density domain wall sample [sample (C)] and its current-mapping analysis. (a) Schematic of the photo-assisted c-AFM setup, illustrating how the domain wall current (DWC) was recorded under controlled illumination conditions using LEDs with different emission wavelengths. (b) Piezoresponse force microscopy (PFM) image taken over a 50 µm x 50 µm area, clearly showing the DW boundaries in sample (C). (c) to (e): c-AFM maps recorded over the same area as in (b), displaying the DWC for 3 different illumination settings, (c) in darkness, (d) at a 330 nm illumination, and (e) for a 310 nm illumination, respectively. (f) to (h): Histograms of figures (c) to (e), clearly indicating the increased DWC with decreasing wavelength.

Supplementary Material

# Comparative study of photo-induced electronic transport along ferroelectric domain walls in lithium niobate single crystals

L. L. Ding,[1,2,3,4,a)] E. Beyreuther,[2] B. Koppitz,[2] K. Kempf,[2] J. H. Ren,[1,3,4] W. J. Chen,[1,3,4,5] M. Rüsing,[2,6] Y. Zheng,[1,3,4] and L. M. Eng[2,7,a)]

[1]Guangdong Provincial Key Laboratory of Magnetoelectric Physics and Devices, School of Physics, Sun Yat-sen University, Guangzhou 510275, China

[2]Institute of Applied Physics, TUD Dresden University of Technology, 01062 Dresden, Germany

[3]State Key Laboratory of Optoelectronic Materials and Technologies, School of Physics, Sun Yat-sen University, Guangzhou 510275, China

[4]Centre for Physical Mechanics and Biophysics, School of Physics, Sun Yat-sen University, Guangzhou 510275, China

[5]School of Materials, Shenzhen Campus of Sun Yat-sen University, Shenzhen 518107, China

[6]Institute of Photonic Quantum Systems, Integrated Quantum Optics, Paderborn University, 33098 Paderborn, Germany

[7]ct.qmat: Dresden-Würzburg Cluster of Excellence – EXC 2147, TU Dresden, 01062 Dresden, Germany

[a)]Authors to whom correspondence should be addressed: lukas.eng@tu-dresden.de, dinglli@mail2.sysu.edu.cn

**Supplement S1:** The current-voltage curves of samples (B) and (C) before the DW-conductivity-enhancement process.

First, we completed the UV-assisted poling process and got the initial hexagonal domain of samples (B) and (C). Next, both sides of our samples were covered with 5-nm-thick chromium electrodes (1x1 mm$^2$). Then, we contacted the electrodes to an electrometer measuring the current-voltage curves of these two samples (B) and (C). Here we show the original I-V curves belonging to (B) and (C), respectively.

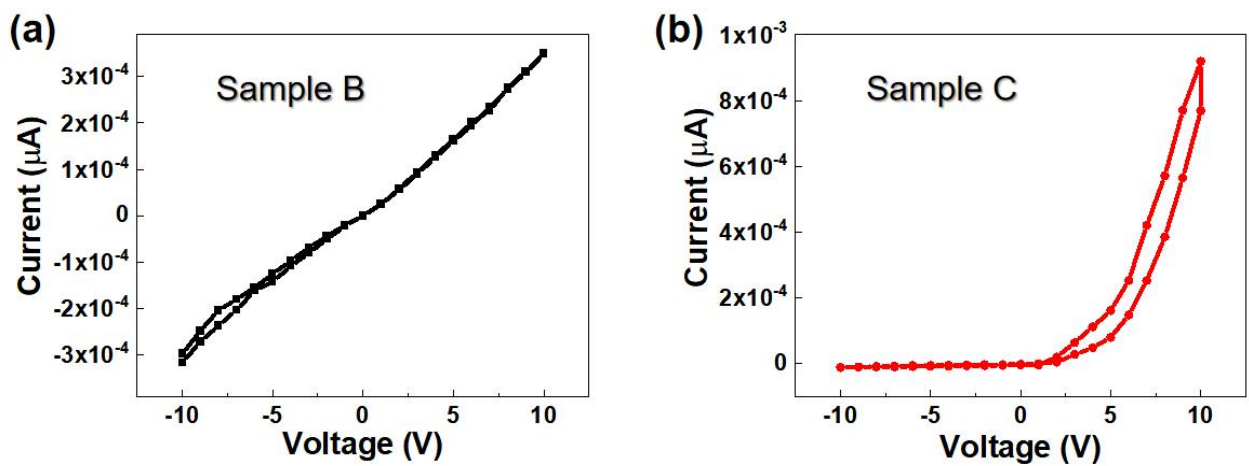

Fig. S1. The current-voltage curves of samples (B) and (C) before the DW-conductivity-enhancement process. (a) and (b) belong to samples (B) and (C), respectively. The original I-V curve of sample (B) is similar to an Ohmic behavior, whereas the original current of sample (C) shows a diode-like tendency.

**Supplement S2:** Experimental details of the SHG method.

In this work, we applied the second harmonic generation (SHG) method to characterize the domain structures of lithium niobate single crystals. The SHG measurements were performed with a commercial laser scanning microscope (Zeiss LSM980MP) using a tunable Ti:Sa laser (Spectra Physics InSight X3, 690–1300 nm, 2 W). For the measurement a laser output power of 40 mW and a wavelength of 900 nm was used. The SH signal is collected in reflection direction using a focusing numerical aperture of NA=0.8. The sample is scanned in the x-y plane using a galvanometer mirror array. By moving the focal point into the sample at different depths, a complete 3D image of the sample can be generated. The mechanism of 3D domain wall visualization in ferroelectric materials with Cherenkov SHG is described in detail in Kämpfe et al. [S1].

**Supplement S3:** Calculation of transmittance of chromium electrode at different wavelengths.

Since in our experiments, we chose chromium as the electrode material for further photo-induced conductivity measurement, a nearly-flat transmittance in the relevant wavelength range should be guaranteed. Here, we calculate the transmittance of chromium electrode under different wavelengths which were used in this work. We could see the transmittance of different wavelengths does not change too much.

| Wavelength | 500 nm | 400 nm | 350 nm | 320 nm | 310 nm |
|---|---|---|---|---|---|
| Transmittance | 0.66015 | 0.63924 | 0.62479 | 0.61571 | 0.61556 |

Table S1. The relatively flat transmittance of chromium electrode (5 nm) under different wavelength illumination conditions. Calculated via the website:

https://refractiveindex.info/?shelf=main&book=Cr&page=Johnson.

**Supplement S4:** Close-up views of electrical transport behaviors in the three samples under different illumination wavelengths.

In order to better compare the electrical transport property of these three samples under different illumination wavelengths, we zoom in some current details dependent on voltage and time under different illumination conditions. Thus, the photo-induced conductivity changes of these three samples could be seen much more clearly.

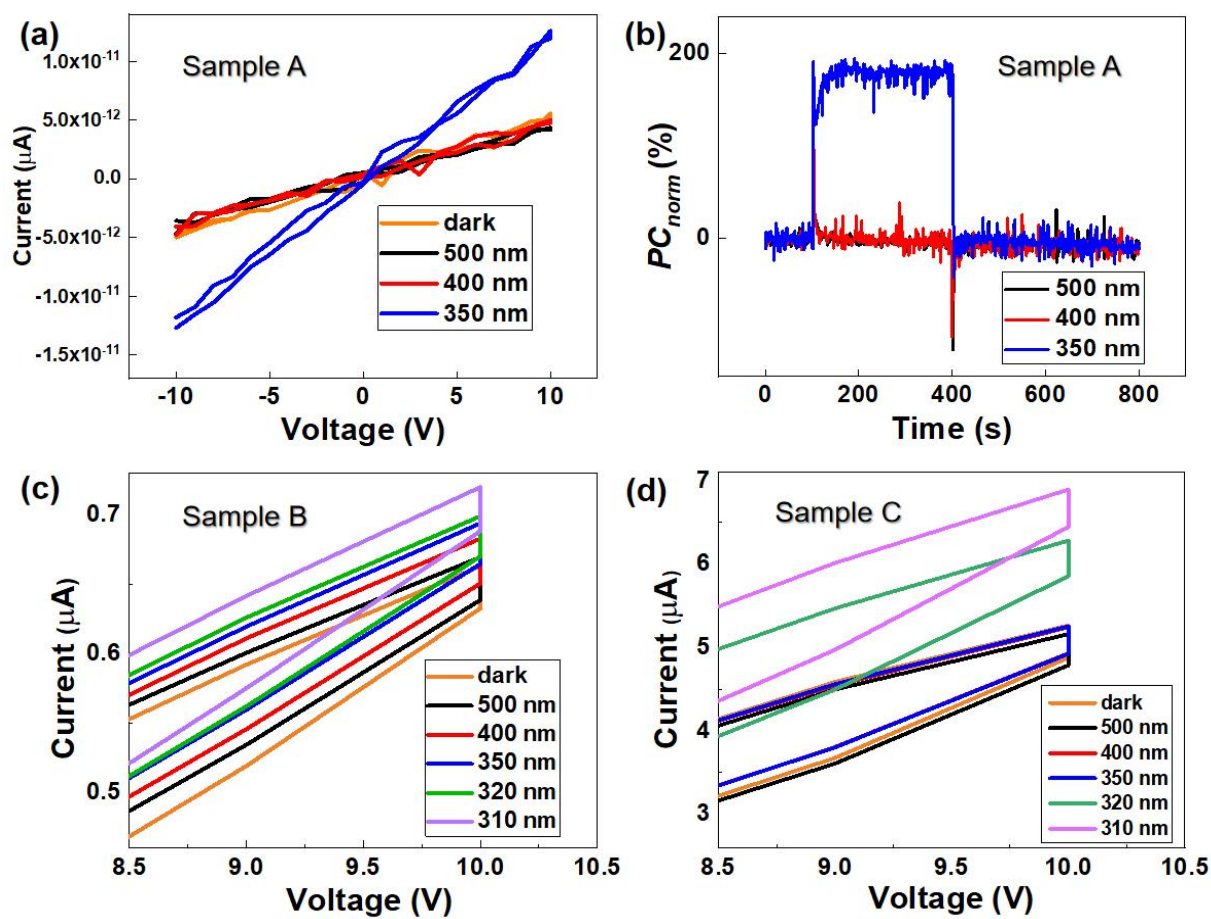

Fig. S2. The zoomed-in I-V curve (in darkness and under the illumination wavelengths of 500 nm, 400 nm, 350 nm) and the photocurrent transients (under the wavelengths, i.e., 500 nm, 400 nm, 350 nm) of sample (A) are shown in (a) and (b). The enlarge I-V curves of samples (B) and (C) are shown in (c) and (d), the voltage varies from 8.5 V to 10 V.

**Supplement S5:** The percentage of the maximum-current increase under different illumination wavelengths.

We summarize the maximum current in darkness and under different illumination wavelengths, and calculate the percentage the maximum current's increase for these three samples. We find that all of three samples reach the maximum current value at the wavelength of 310 nm.

| Maximum current (A) under +10 V | Sample A | Sample B | Sample C |
|---|---|---|---|
| in darkness ($I_{dark}$) | $5.68 \times 10^{-12}$ | $6.63 \times 10^{-7}$ | $5.26 \times 10^{-6}$ |
| 310 nm ($I_{max}$) | $2.21 \times 10^{-9}$ | $7.26 \times 10^{-7}$ | $6.90 \times 10^{-6}$ |
| $I_{ratio}=I_{max}/I_{dark}$ | 389.1 | 1.2 | 1.3 |
| $PC=(I_{ratio}-1)\times 100\%$ | 38810% | 20% | 30% |

Table S2. The maximum current ratio and photoconductivity percentage of the three samples in darkness or under different illuminations, all three samples exhibit the maximum current under the illumination wavelength of 310 nm. $I_{ratio}$ is equal to the ratio of $I_{max}$ and $I_{dark}$. *PC* is the percentage of current increase on basis of $I_{ratio}$.

**Supplement S6:** The raw-data current changes over time in darkness and under different illumination conditions for the three samples.

Here we show the temporal behavior of the total current of three samples under different illumination wavelengths.

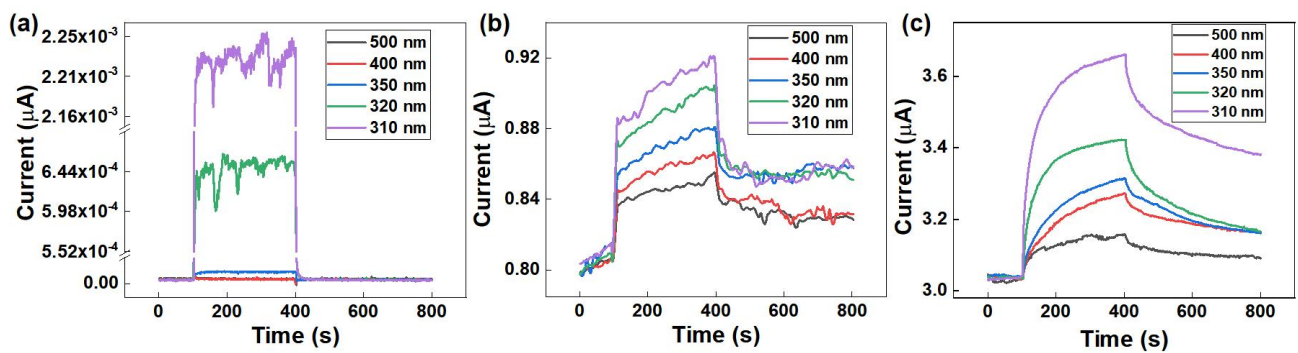

Fig. S3. The original (non-normalized) current transient plots that vary with different illumination conditions. (a), (b) and (c) refer to samples (A), (B) and (C), respectively.

**Supplement S7:** The photoconductance relaxation behavior upon switching the photoexcitation on and off.

For obtaining an idea of characteristic time scales for the photoconductivity reaching saturation, functions of the form

$$I_{norm}(t) = A \cdot exp[-(t/\tau)^\beta] + I_{norm,\infty}$$

were fitted to the measured values of the photoconductivity normalized to the value in darkness, namely $I_{norm}(t)$. This model was employed since it is well known to be suitable for fitting the time evolution of the photoconductivity of oxidic materials, such as in ref. [S2]. In the equation above, $A$ denotes the scaling factor for the time-dependent term leading to the deviation from the saturation value of the photoconductivity $I_{norm,\infty}$, which is reached at large times $t$. $\tau$ is the characteristic time scale for the saturation process to take place, whereas $\beta$ is the exponent which determines by how much the exponential function will be stretched, such that for 0 < $\beta$ < 1, the function will be stretched, and for $\beta$ > 1, it will be compressed in *t*-direction.

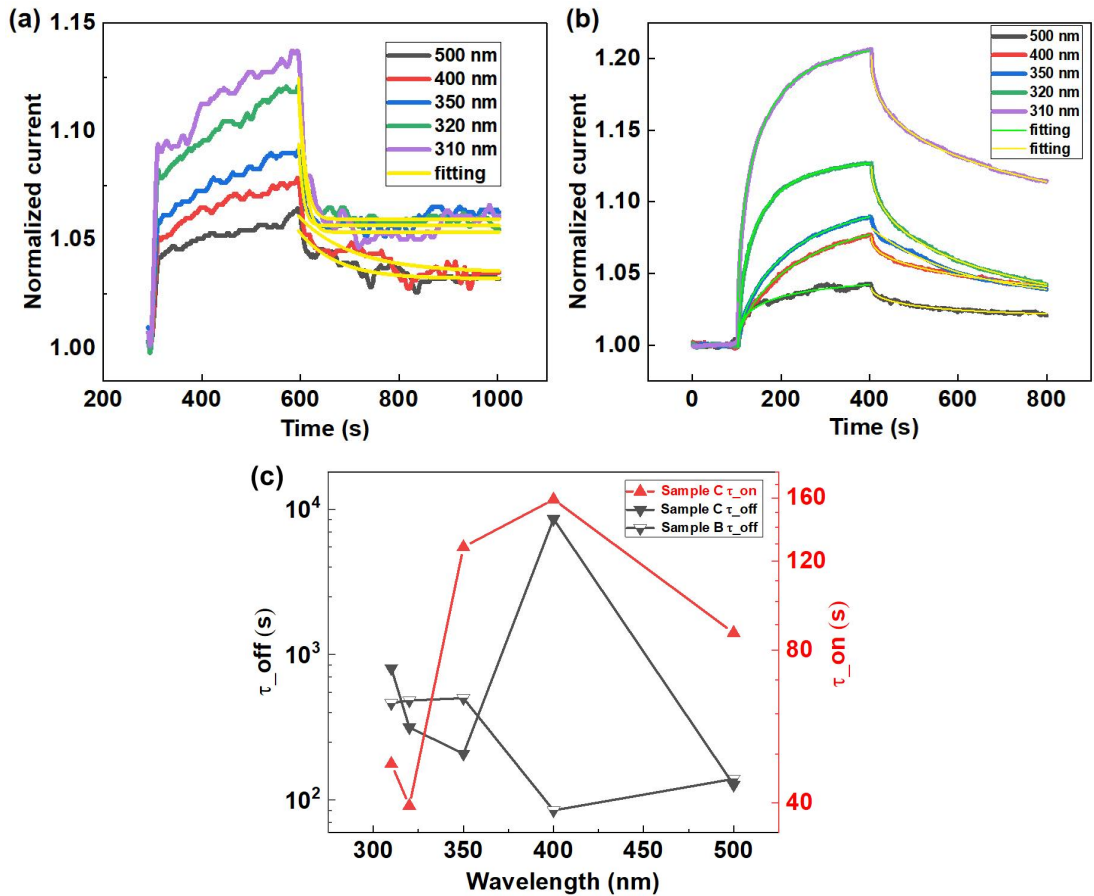

Fig. S4. (a) and (b) are the fitting results about the temporal behaviors of the current in samples (B) and (C), (c) is the summary about the time constant $\tau$ plots of samples (B) and (C).

**Supplement S8:** Comparison of the activation energies calculated from two samples with different domain wall densities under different illumination wavelengths.

| Illumination | Single-domain sample | Multi-domain sample |
|---|---|---|
| darkness | 61.3 | 96.5 |
| 500 nm | 62.2 | 111.8 |
| 400 nm | 62.5 | 106.1 |
| 300 nm | 61.6 | 172.7 |

Table S3. Activation energies in meV, extracted from temperature dependent current measurements of two samples (the first sample with only one hexagonal shaped domain inside, another one with high density domain walls) with different density of DWs under three illumination conditions, exemplarily evaluated at voltages U = +10 V.